# "Parity Inversion And Different Properties Of $_4$Be$^{11}$ Halo Nuclei"

Praveen Chandra Srivastava*and Indira Mehrotra
Department of Physics University of Allahabad, Allahabad, India
*E-mail: chandrapraveen1@rediffmail.com

The recent developments of radioactive nuclear beams has opened up the possibility of exploring a wide variety of nuclei far from the valley of beta-stability. Of late several theoretical studies have come up in the past for describing nuclei in these exotic region. Extension of traditional shell model techniques to explain the exotic feature in the structure of these nuclei such as, formation of neutron halo large r.m.s radii, soft dipole resonances etc have confirmed the necessity to modify the single particle shell model potential when dealing with nuclei far from stability. The ground state nuclear structure properties of $_4$Be$^{11}$ have been calculated in the framework of shell model using analytically soluble mean field potential given by Ginocchio. The potential is highly versatile in nature and depends on four parameters, which define its depth, range and shape. Potential parameters, which generate highly diffuse shape, account for the small binding and halo structure of the valence neutron in $_4$Be$^{11}$. The problem of ground state parity inversion is also addressed.

In the present work we have studied the ground state nuclear structure properties of neutron rich nucleus $_4$Be$^{11}$ using a flexible mean field potential given by Ginnocchi [1]. The naive picture of ground state of $_4$Be$^{11}$ will emerge as neutrons filling up of $0s_{1/2}$, $0p_{3/2}$ state completely and last neutron in $op_{1/2}$ state resulting in ground state $J^\Pi$ as as $(1/2)^-$ whereas experimentally it is known to be $(1/2)^+$.

There have been many theoretical attempts [2-5] to address the problem of parity inversion of ground state and first exited state of $_4$Be$^{11}$ and evaluating extension of its neutron halo, but there is no consistent theory so far. The underline similarty of all these calculations has been to renormalize the average one body potential due to one effect or the other. It has been already observed [6] that the ground state density distribution of $_4$Be$^{11}$ is more or less spherical in shape, which has also been confirmed theoretically [7]. In view of this we have studied the density profile and the halo structure of $_4$Be$^{11}$ by approximating the mean field potential with Ginocchio potential . Earlier work done at analyzing the local equivalent of the full Hartree-Fock potential in nuclei with large neutron excess suggest that the potential develops an increasingly large diffusion as neutron excess grows [8-9]. Clearly this feature can not be accounted for by harmonic oscillator basis as is typically used for stable nuclei. Ginocchio potential has the advantage of having several parameters describing different shapes and depths. For certain combination of these parameters it behaves like usual Woods-Saxon potential with small diffuseness as appropriate for stable nuclei, whereas for certain other combination of parameters it develops large diffuseness appropriate for description of nuclei with large neutron excess.

In the present work we have represented the mean field potential by the Ginocchio potential because of its versatility as described above. It involves four parameters: scaling parameter (s), shape parameter ($\lambda$), effective mass parameter (a) and depth parameter ($\nu$). The mathematical form is given by:

$$W_1(R) = V\, U_1(r) \tag{1}$$

where

$$U_1(r) = \frac{v_1(r)}{\mu(r)} + \frac{c_1(r)}{\mu(r)} \tag{2}$$

where $\mu(r) = 1 - a + ay^2$.

The first potential is

$$v_1(r) = -\lambda^2 v_1 (v_1 + 1)(1 - y^2) + \left(\frac{1-\lambda^2}{4}\right)(1-y^2)(2 - (7-\lambda^2)y^2 + 5(1-\lambda^2)y^4)$$

$$+ \frac{a}{\mu^2}(1-y^2)(1-(1-\lambda^2)y^2)[1 - a + (a(4-3\lambda^2) - 3(2-\lambda^2))y^2$$

$$+ 5(1-\lambda^2)(1-a)y^4 + 2a(1-\lambda^2)y^6] \tag{3}$$

The remaining potential is

$$C_1(r) = \left[\left(\alpha_1^2 - \frac{1}{4}\right)\left(\frac{1-y^2}{y^2}\right)(1 + (\lambda^2 - 1)y^2) - \frac{l(l+1)}{r^2}\right] \tag{4}$$

where r is a dimensionless constant

$$r = sR \tag{5}$$

R is the radial distance and s is the scaling parameter,

$$s = \left[\frac{2mV}{\hbar}\right]^{1/2} \tag{6}$$

Thus in general each nl shell of Ginocchio potential may have its own set of parameters and they may also be different for neutrons and protons. For any choice of parameters potential is analytically soluble and the eigen function $\psi_{nl}$ is given by:

$$\psi_{nl}(R) = (s^{3/2}/r)\phi_{nl}(r) \tag{7}$$

and $\phi_{nl}(r)$ is

$$\phi_{nl}(r) = N(n,l)\frac{[1+\lambda^2(1-a)-(1-\lambda^2(1-a))x]^{1/2}}{[2(1+\lambda^2-(1-\lambda^2)x)]^{1/4}} x \left(\frac{1+x}{2}\right)^{\beta_{nl}/2} \left(\frac{1-x}{2}\right)^{2\alpha_1+1/4} P_n^{(\alpha_1,\beta_{nl})}(x)$$

(8)

The polynomials P are Jacobi's polynomials in which the rank is determined by the radial quantum number n.

$$\text{and } N(n,l) = \left[\frac{2\lambda^2 n!\Gamma(\alpha_1+\beta_{nl}+n+1)\beta_{nl}(\alpha_1+\beta_{nl}+2n+1)}{\Gamma(\alpha_1+n+1)\Gamma(\beta_{nl}+n+1)(\beta_{nl}\lambda^2(1-a)+\alpha_1+2n+1)}\right]^{1/2}$$

(9)

$$\beta_{nl} = \frac{[(2n+\alpha_1+1)^2+\lambda^2(1-a)((v_1+\frac{1}{2})^2-(2n+\alpha_1+1)^2)]^{1/2}-(2n+\alpha_1+1)}{\lambda^2(1-a)}$$

(10)

The eigen energy $E_{nl}$ is

$$\xi_{nl} = -\lambda^4 \beta_{nl}^2 \qquad (11)$$

$$E_{nl} = \xi_{nl} V \qquad (12)$$

We have fixed up values for potential parameters for core nucleus to give correct ground state energy of $Be^{10}$ nucleus. The scale parameter s is chosen to fit the experimental root mean square charge radius. We vary strength parameter $v$ to reproduce experimental separation energy of 0.501Mev for the $1s_{1/2}$ orbit. Variation of $0p_{1/2}$ and $1s_{1/2}$ singe particle energies with parameter $v$ and a are shown in figure1. In table 1 we give value of the parameters for each state. For $2<v<2.2$, $0p_{1/2}$ state lies above $1s_{1/2}$ state .The corresponding $0p_{3/2}$-$0p_{1/2}$ splitting energies is 6.0 MeV indicative of a large spin-orbit interaction. The single particle potential for different states are shown in figure 2. The calculated density profile are shown in Figure 3. The r.m.s. radii are given in table 2.

The results of our calculations suggest that if each shell in the $_4Be^{11}$ nucleus has it own parameters and the potential describing the last neutron has high diffusivity, it is possible to give a shell model description to the density profile and the r.m.s radius. The versatile Ginocchio potential is a good option to represent the mean field potential. We also tried alternative configuration of last neutron of $Be^{11}$. However the centrifugal force in the $0p_{1/2}$ orbit prevents both of tail of halo neutron under predicts density in the tail region. Strong spin orbit interaction indicated by large difference in the value of $\nu$ for the $0p_{3/2}$ and $0p_{1/2}$ sate can bring $1s_{1/2}$ sate below $0p_{1/2}$ state. Thus the results of our calculation indicate that by suitable choice of parameters, the ground state nuclear structure properties can be accounted for by the suitable choice of parameters, the advantage of Ginocchio potential over harmonic oscillator and Woods Saxon potential are it is (i) analytically solvable, (ii) highly versatile, and (iii) represent correct diffuseness.

Table 1.

$\nu$ - values of different single particle states

$\lambda = 3.3286$, s=0.10 and a=0.0

| State | $\nu$ | Energy (MeV) |
|---|---|---|
| $0s_{1/2}$ | 6.1 | -7.16 |
| $0p_{3/2}$ | 6.2 | -6.05 |
| $0p_{1/2}$ | 2.0 | - 0.01 |
| $1s_{1/2}$ | 6.1 | -0.50 |

Table 2.

r.m.s. radii of different states

| Nucleus | Configurations | r.m.s.radius (fm) | |
|---|---|---|---|
| | | Cal. | Expt. |
| $_4B0^{10}$ | | 2.37 | 2.46±0.03 |
| $_4Be^{11}$ | $1s_{1/2}$ | 6.39 | |
| | $0p1/2$ | 5.87 | |

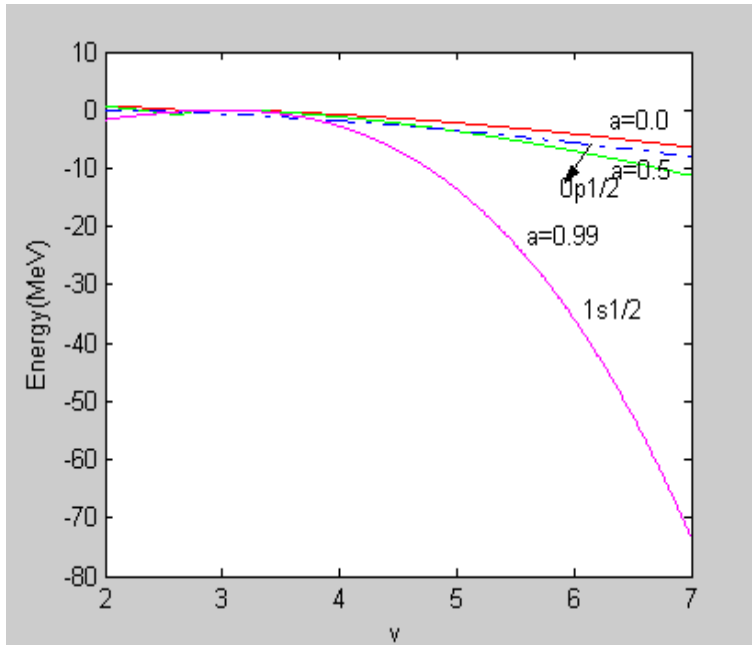

Fig1. Variation of single particle energies with $v$. $0p_{1/2}$ state is plotted for a=0.0, whereas 1s1/2 state is plotted for three different values of a.

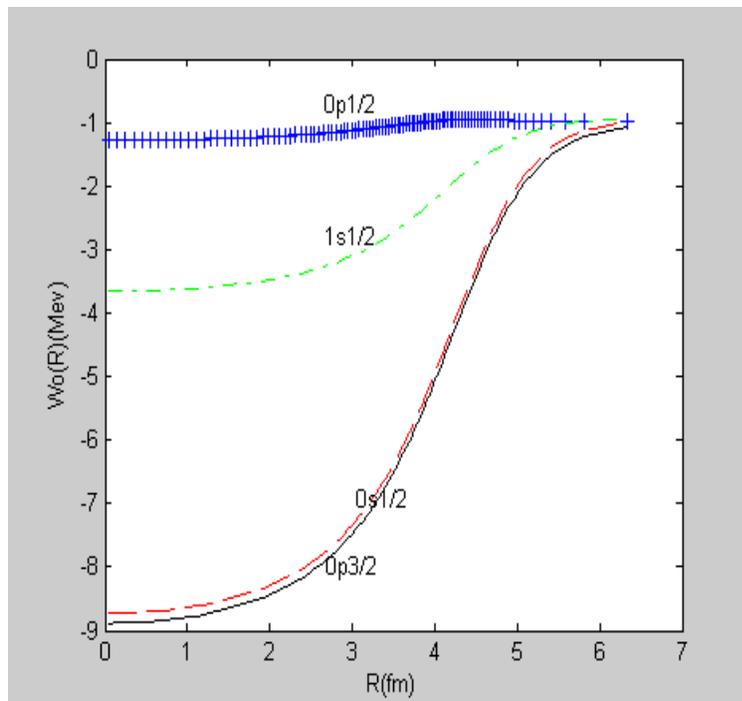

Fig2-Potential for different states with parameters given in table 1 plotted as a function of radius R.

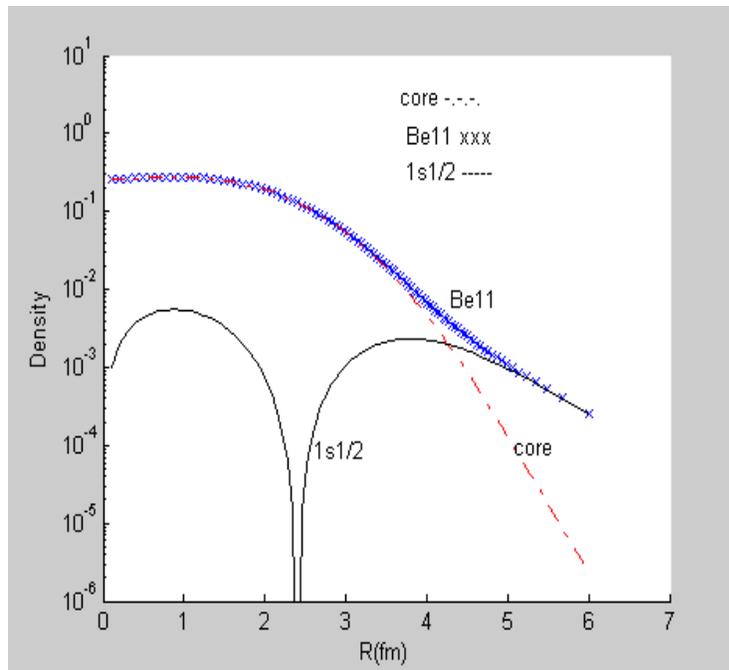

Fig3-.Density profile for core, $1s_{1/2}$ and for $_4Be^{11}$.

**References:**

[1] J. N. Ginocchio, Ann. Phys. 159 (1985), 467.

[2] G. F. Bertsch, B. A. Brown, and H. Sagawa, Phys. Rev C 39 (1989) 1156.

[3] K. Riisager, A. S. Jenson, and P. Moller Nucl. Phys A54 (1992) 393.

[4] R. Bhattacharya and K. Krishan, Phys.Rev. C 56 (1997) 212.

[5] I. Hammoto, H. Sagawa, X. Z. Zhang, Nucl. phys. A 648 (1999)203.

[6] H. Sagawa and K. Yazaki, Phys. Lett. B 149 (1990) 244.

[7] H. Sagawa, Phys. Lett. B 286 (1992) 7.

[8] S. Hanna, K. R. Harris, and  J. W. Olness, Phys. Rev. C 3 (1971) 2198.

[9] F. Ajzenberg - Selove and C. L. Bush, Nucl. Phys. A336 (1980) 1.

[10] M. Fukuda, T. Ichihara etal., Phy. Lett. B 268 (1991) 339.